# Efficient characterization of blinking quantum emitters from scarce data sets via machine learning

G. Landry and C. Bradac*

*Trent University, Department of Physics & Astronomy, 1600 West Bank Drive, Peterborough, ON, K9L 0G2, Canada.*

*Corresponding author, e-mail: carlobradac@trentu.ca

**Abstract**
Single photon emitters are core building blocks of quantum technologies, with established and emerging applications ranging from quantum computing and communication to metrology and sensing. Regardless of their nature, quantum emitters universally display fluorescence intermittency or photoblinking: interaction with the environment can cause the emitters to undergo quantum jumps between on and off states that correlate with higher and lower photoemission events, respectively. Understanding and quantifying the mechanism and dynamics of photoblinking is important for both fundamental and practical reasons. However, the analysis of blinking time traces is often afflicted by data scarcity. Blinking emitters can photo-bleach and cease to fluoresce over time scales that are too short for their photodynamics to be captured by traditional statistical methods. Here, we demonstrate two approaches based on machine learning that directly address this problem. We present a multi-feature regression algorithm and a genetic algorithm that allow for the extraction of blinking on/off switching rates with $\geq$85% accuracy, and with $\geq$10× less data and $\geq$20× higher precision than traditional methods based on statistical inference. Our algorithms effectively extend the range of surveyable blinking systems and trapping dynamics to those that would otherwise be considered too short-lived to be investigated. They are therefore a powerful tool to help gain a better understanding of the physical mechanism of photoblinking, with practical benefits for applications based on quantum emitters that rely on either mitigating or harnessing the phenomenon.

## 1. Introduction

Fluorescence intermittency, also referred to as telegraph noise or blinking, is a phenomenon common to many photon-emitting (quantum) systems, especially when the linear size of the objects or host materials reaches the nanoscale. Affected systems include organic molecules and proteins,[1–3] polymers,[4] quantum dots,[5–7] and a large variety of nanostructured materials.[8–14] Under continuous optical excitation, these photo-intermittent systems undergo stochastic switching between two or more states, each corresponding to a distinct recombination/decay path for the excited electrons (and holes) involved (cf. §§ 2.1–2). As a result, the emitters display different and characteristic step-like fluorescence intensities as they transition between on/off states with temporal dynamics generally in the ~ms–s range (Figure 1).

Generalized models attempting to capture the origin of photoluminescence intermittency and quantify the switching rates have been proposed.[12,15,16] They mostly revolve around the idea that, upon excitation, electrons and/or holes become involved in trapping mechanisms with characteristic lifetimes. The emitter-to-emitter and system-to-system variabilities[7,14,17–20] suggest the existence of system-specific trapping mechanisms that are relevant as they can reveal important information about the nature of the emitters and their interaction with the local environment. This goes beyond mere fundamental interests: understanding and quantifying the blinking photodynamics have practical value, for they can lead to either mitigate its occurrence,[21–25] when undesired, or harness it for applications that rely on it, such as in stochastic superresolution microscopy.[18–20,26,27]

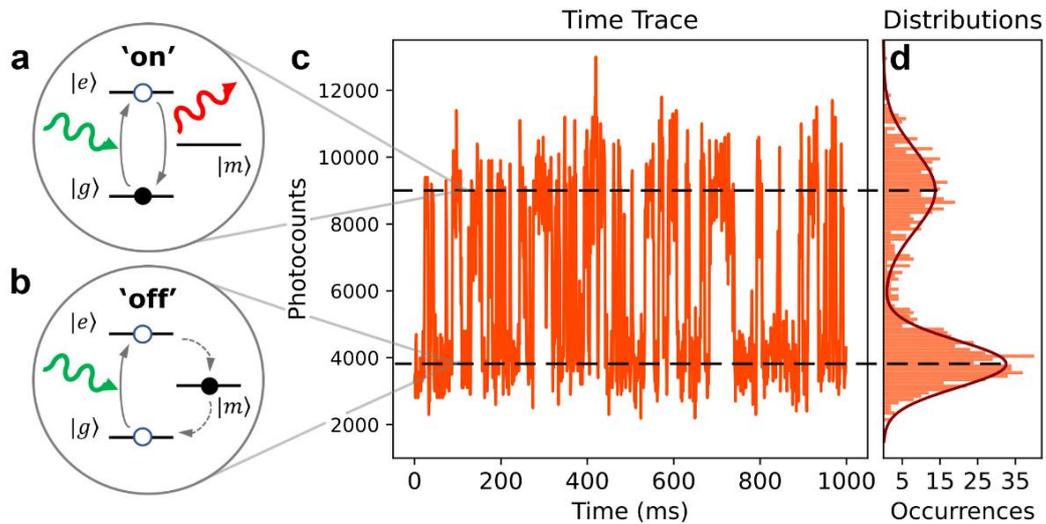

**Figure 1.** Fluorescence intermittency. **a, b)** Schematic representation of the 'on' (a) and 'off' (b) states. The on state corresponds to the emitter continuously cycling between the ground, $|g\rangle$, and excited state, $|e\rangle$, via optical excitation and radiative decay. The off state corresponds to the emitter 'losing' an electron to a relatively long-lived metastable/trap state, $|m\rangle$. **c)** Time trace showing the difference in fluorescence photocounts between the on/bright and the off/dark states. **d)** Statistics of the on and off states' occurrences.

Regardless of the strategy chosen for extracting blinking statistics from photoluminescence time traces,[16,28,29] one of the main hurdles any approach faces is data scarcity. The average permanency of blinking emitters in a given state generally ranges from milliseconds to seconds and can vary from system to system. Therefore, building up enough statistics to accurately determine their switching rates requires time traces that are tens to several hundreds of seconds long. The ratio between acquisition time and average state lifetime should be, at minimum, ~$10^3$—realistically, ~$10^4$ or even ~$10^5$. This can be problematic, for in many cases blinking emitters can undergo photo-bleaching at shorter time scales.[30,31] Here, we demonstrate two approaches based on machine learning (ML) that address directly this problem and allow for the extraction of blinking rates with at least an order of magnitude less data and higher precision than traditional methods based, e.g., on statistical inference. These ML methods thus expand the range of surveyable blinking systems to those that are relatively short-lived—potentially revealing trapping dynamics that would otherwise be impossible to capture due to insufficient statistics of switching events. To showcase their accuracy, efficiency and precision, we employ these ML strategies to characterize the switching rates of emulated blinking emitters and we compare them to traditional statistical methods. When relevant, we discuss how these ML algorithms can be generalized to characterize more complex dynamics and systems—potentially beyond photoblinking emitters.

## 2. Theoretical framework
The methods we propose, their analysis and benchmarking against other approaches are based on a series of hypotheses and assumptions that we detail below.

*2.1 On/off states and (generalized) quantum jump model*

In this work, our framework for describing the blinking mechanism is the quantum jump model.[31] Blinking emitters are effectively treated as generalized three-level systems that perform stochastic jumps between an *on/bright* and an *off/dark* state, respectively characterized by a high and low recorded intensity in the photoluminescence time trace (Figure 1). In the simplest version of this model, under continuous optical illumination an electron in the system undergoes consecutive excitation and radiative recombination cycles between the ground, $|g\rangle$, and excited level, $|e\rangle$. The system is said to be in the on/bright state, $[|g\rangle \leftrightarrows |e\rangle]_{on}$, and displays a high number of fluorescence photocounts. From the on state the system can transition, with rate $k_{on}$, to the off/dark state, $[|m\rangle]_{off}$. This can be a metastable state, e.g., an acceptor-like state originating from surface or local reconstruction of the crystalline structure surrounding the emitter. In the off state, photoemission ceases until the electron relaxes non-radiatively and transitions back, with an overall rate $k_{off}$, to the on state, resuming the photoemission cycle. Note that the time scales of the radiative recombination and trapping/non-radiative recombination mechanism(s) can be vastly different: ~$10^{-9}$–$10^{-7}$ s for the radiative relaxation of the electron in the on state, $|g\rangle \leftrightarrows |e\rangle$, vs. ~$10^{-3}$–$10^{0}$ s for the trapping/non-radiative recombination process, $[|g\rangle \leftrightarrows |e\rangle]_{on} \leftrightarrows [|m\rangle]_{off}$. Specific systems can display more complex dynamics (both for the on and off states), for instance in the case of trapping mechanisms involving multiple recombination channels each with its individual—potentially varying over time—trapping/recombination rate.[15] For these, the simple quantum jump model can be generalized by defining a total non-radiative recombination rate, $k_t(t)$, that accounts for all trapping mechanisms:

$$k_t(t) = \sum_{j=0}^{J} k_j \sigma_j(t) \qquad (1)$$

The terms $\sigma_j(t)$'s are such that $\sigma_j = 1$ when the corresponding $j$-th trapping channel is active and $\sigma_j = 0$ when it is inactive. Equation (1) is general and can describe systems of different dynamics (cf. §§ 2.2, 2.3) as the active-to-passive, $\sigma_j(t) = 1 \to 0$, and passive-to-active, $\sigma_j(t) = 0 \to 1$, distributions can be functions varying with time with switching rates $\gamma_j^-$ and $\gamma_j^+$, respectively. The quantity $k_0$ is the background rate and is always active ($\sigma_0 = 1$), even when any other trapping channel is inactive. Under the hypotheses that the electronic excitations and radiative recombinations of rates $k_I$ and $k_r$, respectively, are much faster than the trapping rates $k_j$, i.e., $k_I, k_r \gg k_j \; \forall j$, the steady-state expression for the photoluminescence intensity of a blinking emitter can be written as

$$I(t) = \frac{k_I}{k_I + k_r + k_t(t)} \qquad (2)$$

which captures the photoluminescence jumps observed for blinking emitters. In fact, when every trapping mechanism is inactive, $k_t(t) = 0$, the intensity $I(t)$ is maximum (dependent simply on the excitation and radiative rates $k_I$ and $k_r$), while it lowers when $k_t(t) \neq 0$ (i.e., when at least one trapping channel is active). Additionally, with respect to an arbitrary threshold separating the on and off states such that the emitter is on when $I(t) > I_{th}$ and off when $I(t) < I_{th}$, we can express the on/off condition with respect to a threshold rate $k_{th}$. Specifically, if $k_t(t) \leq k_{th}$, the emitter is on, and otherwise it is off.

As a final note, while it is generally true that trapping processes are much slower (up to a few orders of magnitude) than electronic transitions ($k_I, k_r \gg k_j \; \forall j$), rapid trapping events can occur, indicating that individual systems can display a vast range of diverse and complex dynamics deviating from the simplest of models.[18]

*2.2 Analytical form of the on/off distributions*

The distribution of the on and off events in blinking emitters is, generally, system-dependent (and in some cases also emitter-dependent). Blinking emitters are therefore usually organized in two main categories:[32] *i)* systems for which the distribution of the on and off times is exponential or near exponential, $P(\tau_{on/off}) \propto e^{-t/\tau_{on/off}}$ and *ii)* systems for which the distribution of the blinking events follows an inverse power law, $P(\tau_{on/off}) \propto \tau_{on/off}^{-m_{on/off}}$. The first category includes, e.g., single molecules,[1,33] polymers[4] and quantum emitters in solid-state hosts,[13,14] while the second includes mainly quantum dots.[15,28,34] This warrants a couple of observations. We note, firstly, that this classification is not absolute as any of these systems can deviate from the 'nominal' behavior—indicating that the trapping mechanisms at play for a given blinking emitter can be very nuanced and highly dependent on its local environment. We remark, also, that mathematically the dynamics for both systems of type (*i*) and (*ii*) are captured by the general equation (1), once the number of trapping mechanism(s), as well as their rates and time distributions are set. Broadly speaking, the probability density of the on/off time occurrences following an exponential decay is indicative of a single trapping mechanism with individual on and off dynamics characterized by corresponding average lifetimes $\langle \tau_{on} \rangle = \langle 1/k_{on} \rangle$ and $\langle \tau_{off} \rangle = \langle 1/k_{off} \rangle$, respectively. Conversely, an inverse power law is indicative of a distribution of trapping dynamics consisting of either *a)* an individual trapping mechanism that undergoes fluctuations over time and thus displays a distribution of on/off rates or *b)* a large set of co-existing trapping mechanisms each with its own rate. A combination of scenarios (*a*) and (*b*) is also possible and is captured, again, by the general equation (1). Note that a set of exponential distributions—either discrete but sufficiently dense and weighted, or nearly-continuous—can reproduce the inverse power law behavior. Detailed analyses involving changes in the binning times can reveal the multi-exponential vs. inverse power law distribution of $P(\tau_{on/off})$, as $\langle \tau_{on/off} \rangle$ is respectively independent of the integration window for the former, and dependent for the latter.[31]

*2.3 Synthetic data*

To demonstrate the accuracy, efficiency and precision in extracting the on/off rates of blinking emitters for the ML methods we hereby propose, we use synthetic data that is generated computationally. This allows us to specify—arbitrarily—the on and off transition rates of the blinking emitters and their corresponding probability densities $P(\tau_{on})$ and $P(\tau_{off})$. It also allows us to fully control other (simulated) experimental parameters such as total duration, thresholds and signal-to-noise ratio of the data. Since the emulated blinking systems are known with absolute certainty, we can use the synthetic fluorescence data to benchmark the performance of our methods against those based on traditional statistical inference, e.g., Levenberg-Marquardt (L-M) or damped least-squares fit of $P(\tau_{on/off})$.[35,36]

We make two hypotheses for the purpose of our demonstrations.

Hypothesis *I*: we emulate blinking emitters with on and off states characterized by survival, $N_{on/off}(\tau)$, and switching probabilities, $\bar{N}_{on/off}(\tau)$, respectively given by:

$$N_{on/off}(\tau) = e^{-\tau/\tau_{on/off}} \quad (3)$$

$$\bar{N}_{on/off}(\tau) = 1 - e^{-\tau/\tau_{on/off}} \quad (4)$$

where $\tau_{on/off}$ is the average lifetime of the state. Note that equations (3) and (4) imply the following. The quantity $N_{on/off}(\tau) = e^{-\tau/\tau_{on/off}}$ is effectively the probability that an emitter starting in the on/off state remains in the same on/off state over time. The probability density $P(\tau_{on/off})$ is thus obtained from the time derivative of $N_{on/off}$, i.e., $P(\tau_{on/off}) = -dN_{on/off}/d\tau = -(1/\tau_{on/off})e^{-\tau/\tau_{on/off}}$. The probability of a state to remain populated is independent of the other state. Each transition from one state to another begins a new Poisson process with the same per-state probability distribution $P(\tau_{on/off})$; that is, the rate of transition, and thus the average lifetime of the state, is independent of the total age of the state and the order of the activation. Once the system has changed state it 'forgets' everything about the previous state.

Hypothesis $II$: for simplicity, the on and off states in the emulated photoluminescence time traces are well separated. There is no overlap between the noise of the on and off photocounts—hence a threshold separating the states can be established unambiguously. Figure 2 shows an example of real data (fig. 2a) that is directly comparable with an equivalent synthetic set of blinking data emulated computationally (fig. 2b).

When appropriate, the applicability of our ML methods to cases beyond the quantum jump framework set by hypotheses $I$ and $II$ are presented in the relevant, specific subsections (cf. §§ 3.1, 3.2).

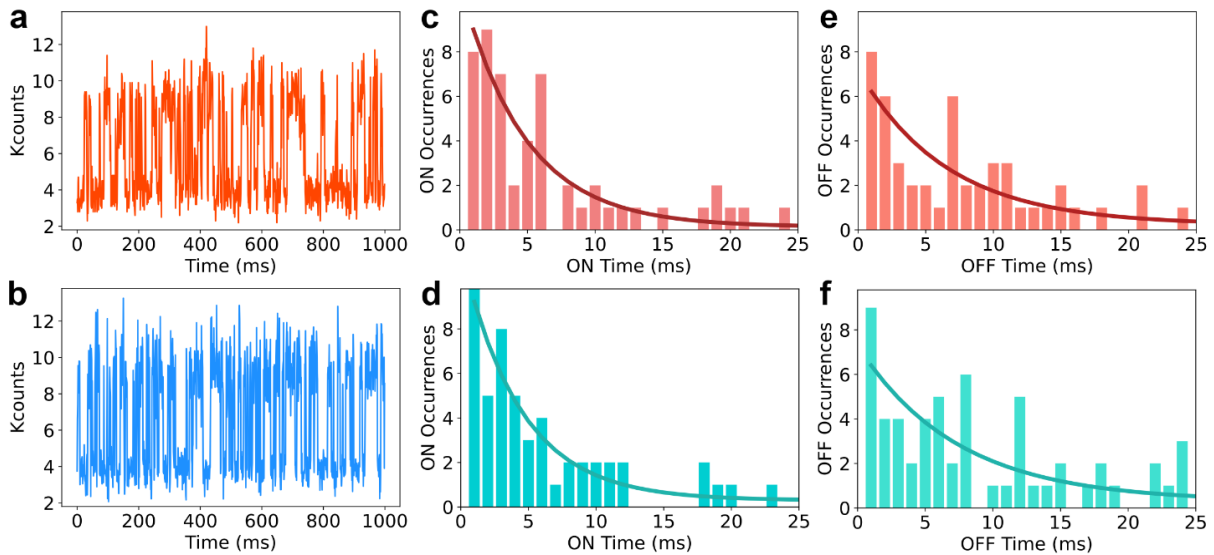

**Figure 2.** Time traces and statistics for real and synthetic blinking emitters. **a, b)** Experimental time trace measured for a blinking nitrogen-vacancy (NV) center in a 5-nm nanodiamond host (a), and simulated time trace for a synthetic blinking center (b). **c–f)** Histograms of the 'on'-state (c, e) and 'off'-state (d, f) average lifetimes for the real and synthetic blinking emitter, respectively. The time constants measured for the real NV center are $\langle \tau_{on} \rangle = 4.8$ ms (c) and $\langle \tau_{off} \rangle = 6.7$ ms (e). These values are used to create the synthetic data in (b) that, in turn, produces the synthetic 'on'- and 'off'-state histograms (d) and (f), counterpart to the respective histograms (c) and (e) of the real NV emitter.

## 3. Results and Discussion

Within the framework we established, characterization of the photoluminescence intermittency of an emitter can be achieved through determination of its on- and off-time probability densities, $P(\tau_{on})$ and $P(\tau_{off})$. Experimentally, this consists in recording from the fluorescence time trace a histogram of occurrences for various time durations (time bins)—i.e., by tallying the frequency with which the blinking emitter is consecutively measured in the on and off states for each time duration (figure 2 c–f). Under hypothesis *I*, the blinking photokinetics in the synthetic data (cf. § 2.3) is governed by one rate constant for the on state and one for the off state, following decaying exponential probability densities $P(\tau_{on})$ and $P(\tau_{off})$ from equations (3) and (4) (note that the measured $P(\tau_{on})$ and $P(\tau_{off})$ are not continuous functions due to the discrete nature of the sampling bin size). Traditional statistical inference methods thus extract the parameters of interest, specifically the average rates at which the emitter switches between states $\langle k_{on} \rangle = \langle 1/\tau_{on} \rangle$ and $\langle k_{off} \rangle = \langle 1/\tau_{off} \rangle$, by fitting the probability densities to $P(\tau_{on/off}) = y_{0,on/off} + A_{on/off} \cdot e^{-t/\tau_{on/off}}$, through standard least square curve fitting (e.g., Levenberg-Marquardt). We note that one of the main issues with this traditional approach is that the density of data recorded—especially for the long-time bins—is usually low. This requires acquiring long fluorescence time traces (~several tens–hundreds of seconds) to capture enough statistics to populate the long-time bins of the histogram and fit the data to the (exponential) model. In the following subsections we show how the ML algorithms we developed can overcome this limitation.

*3.1 Multi-feature regression*

The first ML approach we present to tackle the data-scarcity problem is a simple multi-feature regression (MFR) algorithm. Instead of considering $\tau_{on}$ and $\tau_{off}$ as parameters to be extracted from data fitted to (exponential) models, we define them as functions that depend on several variables or *features*. Specifically,

$$\tau_{on/off,i}(1, x_{i1}, x_{i2}, \dots, x_{in}) = \sum_{j=0}^{n} w_{ij} x_{ij} = w_{i0} + w_{i1} x_{i1} + w_{i2} x_{i2} + \cdots + w_{in} x_{in} \quad (5)$$

Which is a generalization of the slope-intercept linear regression for $n$ features. The time constants $\tau_{on,i}$ and $\tau_{off,i}$ are therefore (weighted) linear combinations of $n$ features: $\boldsymbol{x}_i = [1, x_{i1}, x_{i2}, \dots, x_{in}]$. The quantity $n$ is simply the total number of time-bins in the histogram of occurrences, $x_{ij}$ is the number of occurrences recorded for the *j*-th time-bin and $w_{ij}$ is the corresponding weight (the term $x_{i0} = 1$ with weight $w_{i0}$ accounts for the bias or intercept value of the regression). Note that the MFR algorithm is a *supervised learning* algorithm that requires training with known sets of data. Specifically, the MFR algorithm is given a certain number of *labeled points*, or *training set(s)*, $(\boldsymbol{x}_i, y_i)_i$, where $\boldsymbol{x}_i$ are the time-bin occurrences for the *i*-th data set and $y_i$ is the corresponding *label*, i.e., the value of the observable of interest, $\tau_{on/off,i}$, in our case (the index $i = 1, 2, \dots, N$ runs over the total number $N$ of data sets). For the training sets, both the features $\boldsymbol{x}_i$'s and the corresponding labels $y_i$'s are known. The MFR algorithm uses this information to infer a labeling rule $\boldsymbol{x}_i \to y_i$ that can then be used to determine the value of a label $y_i$ (i.e., $\tau_{on/off,i}$) from an unknown set of features $\boldsymbol{x}_i$ (i.e., occurrences in time-bins) different from the data sets the algorithm was trained on. In our MFR algorithm, determining the labeling rule simply consists in finding the values of the weight coefficients $\boldsymbol{w}_i = [w_{i0}, w_{i1}, \dots, w_{in}]$ that—for the given training sets—minimize a defined *cost function*. For our demonstration, we use a standard least squares error (*LSE*) cost function, or L2-norm loss function, though other cost functions are possible:

$$LSE = \sum_{i=1}^{N} (\boldsymbol{w}_i \boldsymbol{x}_i - \tau_i)^2 \quad (6)$$

Here the subscript $i$ cycles through the different datasets we use for training the algorithm ($i = 1, 2, \ldots, N$) and $\tau_{on/off,i}$ are the target values of the on/off time constants and are known *exactly* for the $N$ sets of training data. We recall that to train our algorithm, but also to test it and compare it to traditional statistical methods, we use synthetic data sets that we emulate computationally and are thus known with absolute accuracy (cf. § 2.3).

We find that the MFR algorithm manifestly outperforms standard statistical inference methods. In our demonstration, we generate several sets of synthetic fluorescence intermittency time traces. Specifically, we produce synthetic time traces with $\langle \tau_{on} \rangle = 15$ ms and $\langle \tau_{off} \rangle = 45$ ms and time bin $1$ ms, chosen arbitrarily. From the fluorescent time traces, we then build the histograms of on and off occurrences per time bin that give the corresponding (discrete) probability densities $P(\tau_{on})$ and $P(\tau_{off})$. The values of $\langle \tau_{on} \rangle$ and $\langle \tau_{off} \rangle$ are extracted using both a Levenberg-Marquardt (L-M) fit of $P(\tau_{on})$ and $P(\tau_{off})$ and the MFR algorithm (which had been previously trained with 20 training sets of synthetic data). The extracted values of $\langle \tau_{on} \rangle$ and $\langle \tau_{off} \rangle$ are then compared to the actual values of $15$ ms and $45$ ms respectively, to establish the accuracy of each approach. Notably, for the same level of accuracy, the MFR algorithm can determine the values of $\tau_{on/off}$ with a ≥10× factor fewer data points than the L-M method. This means that we can extract the same level of information from fluorescence time traces that are tenfold shorter in duration (figures 3a, b). Specifically, after training the MFR algorithm with just only 20 emulated training sets the model can predict the correct value for $\langle \tau_{on} \rangle$ and $\langle \tau_{off} \rangle$ with ≥85% accuracy from a time trace as short as ~0.2 s (i.e., 200 data points) and ~2 s (i.e., 2·10³ data points), respectively. Note that in the case considered here, $\langle \tau_{on} \rangle$ is three times shorter than $\langle \tau_{off} \rangle$, which results in both the L-M fit and the MFR algorithm reaching the same level of accuracy with a shorter time trace (i.e., fewer data points) for $\langle \tau_{on} \rangle$ than for $\langle \tau_{off} \rangle$. The same level of accuracy can be achieved through traditional statistical inference (L-M) methods with much longer acquisition times, ~20–200 s (i.e., 20·10³–20·10⁴ data points). In fact, traditional least square curve fitting methods are incapable of converging at all with a time trace as short as 2 s (2·10³ data points) or less, and if they do, the value of $\langle \tau_{on/off} \rangle$ is only known with an accuracy of ~10% (i.e., with ~90% uncertainty).

These results are highlighted in figures 3a and 3b where the accuracy (measured as the distance between the predicted and the true value) and the corresponding precision (measured as the standard deviation or error of the predicted values) are displayed for $\langle \tau_{on} \rangle$ (fig. 3a) and $\langle \tau_{off} \rangle$ (fig. 3b). To note is that, compared to the traditional L-M fit, the MFR algorithm predicts values of $\langle \tau_{on} \rangle$ and $\langle \tau_{off} \rangle$ that are both more accurate—the markers are closer to the dashed lines corresponding to $\langle \tau_{on} \rangle = 15$ ms and $\langle \tau_{off} \rangle = 45$ ms —and also more precise—the shaded area around the predicted values is smaller. Additionally, as already noted above, the MFR algorithm can predict values of $\langle \tau_{on} \rangle$ and $\langle \tau_{off} \rangle$ more efficiently, i.e., with comparatively fewer data points. Similarly, the heatmaps in figures 3c and 3d offer a visual summary of the higher precision—indicated by smaller numeric error values and by lighter colors—of the MFR and genetic algorithm (GA, cf. § 3.2 below) compared to a standard L-M fit. Note that, in the heatmaps, dark shaded cells with no numeric value indicate that there were not enough data points for the algorithm to determine a value of $\langle \tau_{on} \rangle$ and $\langle \tau_{off} \rangle$; again, the better performance of the MFR and GA algorithms over the standard L-M fit is evident.

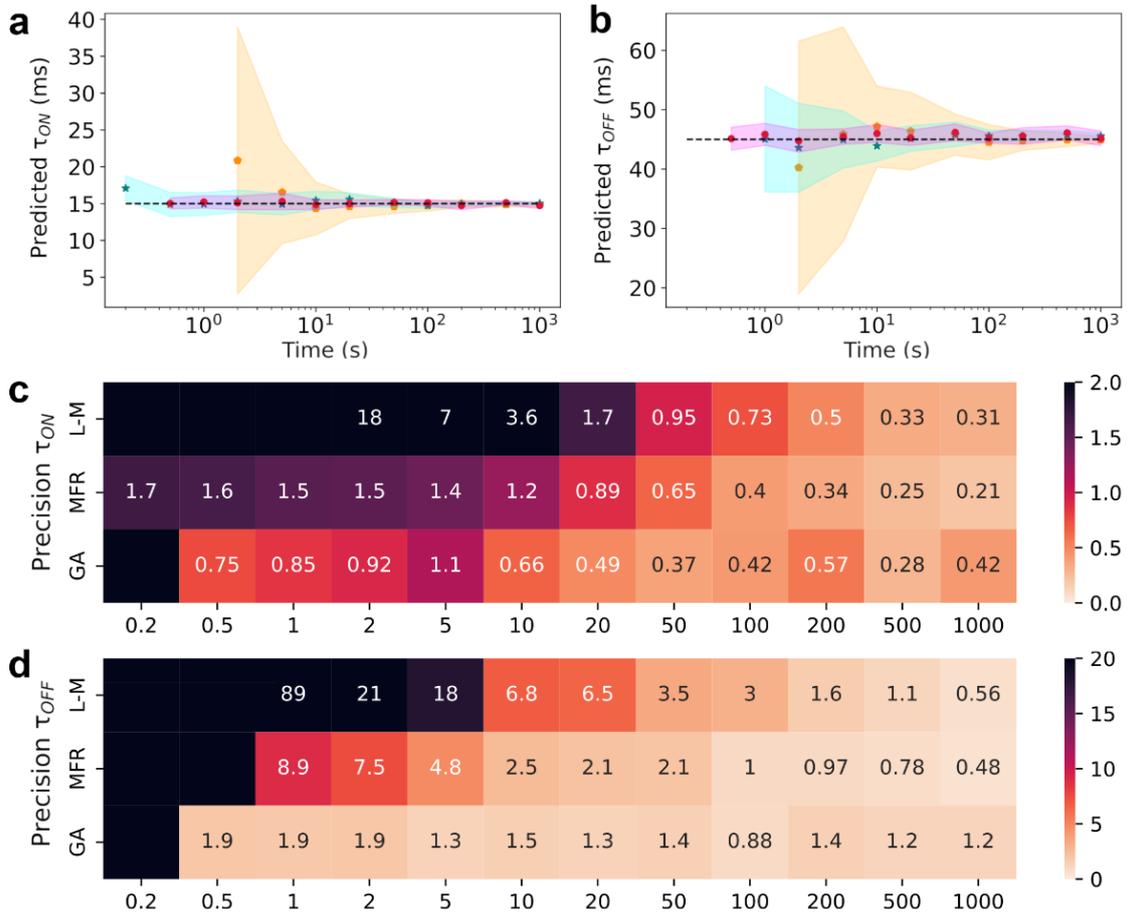

**Figure 3.** Performance comparison of the various models. **a,b)** Nominal value (black, dashed line) of the simulated $\langle \tau_{on} \rangle = 15$ ms (a) and $\langle \tau_{off} \rangle = 45$ ms (b) with the corresponding values and uncertainties predicted by a standard L-M fit (orange diamonds and shading), and by the MFR model (light blue stars and shading) and GA model (red circles and shading). Note how both the MFR and GA models can predict the correct values for $\langle \tau_{on} \rangle$ and $\langle \tau_{off} \rangle$ with comparatively fewer data points (i.e., shorter time traces) than a standard L-M fit and with a smaller uncertainty. **c, d)** Heatmap showing the achieved precision (standard deviation from the nominal values of $\langle \tau_{on} \rangle = 15$ ms (a) and $\langle \tau_{off} \rangle = 45$ ms) for the various models as a function of duration of the time trace (0.2–1·10³ s or 200–1·10⁶ data points). Lighter colors indicate higher precisions, i.e., smaller errors (also shown numerically in the heatmap); if the numeric value is not shown it indicates that the model was not able to converge to a value of $\langle \tau_{on} \rangle$ or $\langle \tau_{off} \rangle$ due to data scarcity. Note how the MFR and GA algorithm outperform the standard L-M fit both in terms of accuracy and precision, as well as a smaller amount of data required to converge.

These results and specifically the difference in performance of the MFR algorithm vs. that of a traditional L-M fit call for a few important observations. Above all, both machine learning and traditional inference-based methods utilize statistical analysis to make determinations from available data. However, they do so differently, even though the difference is conceptual rather than fundamental. Traditional inference-based approaches rely on a proposed model of unknown parameters that are determined using statistical tools from fitting *complete* data sets. The term 'complete' indicates explicitly that the available data ought to be large enough to be fitted by the model within a target level of accuracy. Conversely, machine learning utilizes general-purpose *learning methods* that do not rely on an explicit model for the distribution of the data and make predictions in response to identifying patterns—often from *sparse* data sets. The term 'sparse' indicates that while a single data set might not be large enough to be fitted via a traditional approach, e.g., L-M, many sparse data sets can display enough pattern repetition to allow ML algorithms to correlate the data to specific outcomes. This conceptual difference is sometimes embodied by the expressions *long data* (relevant to inferential statistics) vs. *wide data* (relevant to ML).[37]

Within this context, we ascribe the efficiency of the MFR algorithm to extract emitters' blinking rates to a few key factors. In general—and commonly to many supervised ML strategies—ML algorithms are capable of building strong data-to-outcomes correlations from training data sets of known, previously-acquired data. This usually comes at a high overhead cost: the actual availability of said (wide) data, i.e., of large sets of experimental observations to train the algorithm with. In our case, however, this initial cost is negligible. We use synthetic data from emulated blinking emitters to train our MFR algorithm, which means we can train the algorithm with an arbitrarily large number of data sets. This is possible because we can describe any blinking system from its rate equations with an arbitrary level of control on every parameter: from rate constants and number of trapping mechanisms to noise level and total durations of the fluorescence time traces. This way of 'acquiring' the training sets is noteworthy as it directly overcomes the requirement of gathering, upfront, large sets of experimental training data. It also underscores the vast potential of hybrid approaches[38] where both ML and knowledge-driven modeling are combined to achieve high optimization efficiencies.[39] It however also calls for caution as the MFR algorithm will perform reliably for data within the training parameter space, but might not be generalizable to data outside of it. Nevertheless, traditional L-M fitting approaches rely themselves on pre-assumed models that could be too simple or inaccurate, making that of generalization a problem that is not just one of machine learning approaches alone.

Besides the caveat of generalizability, one of the more specific reasons the MFR algorithm performs better than a standard L-M fit is that it is less reliant on low density of data, especially in the long-time bins of the histogram of occurrences. Having potentially different weights $w_{ij}$ for different time bins $x_{ij}$ in the MFR algorithm directly mitigates the data-scarcity problem, as each bin can be weighted more/less. Notably, this has been shown to be advantageous for other types of vastly different systems and measurements that utilize a similar ML base multi-regression approach—indicating that it is a core trait of the approach.[40]

As a final note, it is worth emphasizing the generality of the MFR algorithm well beyond the framework we set (cf. §§ 2.1–3 and hypotheses *I* and *II*). With reference to hypothesis *I*, we remark that the MFR algorithm is entirely independent of the probability distributions $P(\tau_{on/off})$. Equations (5) and (6) are general and do not rely on a specific analytical expression for $P(\tau_{on/off})$—it be exponential, described by an inverse power law or any combination of these. This follows, again, from a trait common to many machine learning strategies, namely the use of general-purpose learning methods that do not rely on an explicit model. Additionally, with reference to hypothesis *II*, the method can be applied to strategies that do not require the separation of the on and off states by a threshold, including for instance those based on the analysis of the power spectral density (PSD) of fluorescence time traces.[15,29] In fact—whilst requiring to be adapted to specific cases—the MFR algorithm is universally applicable, only requiring for the measured signal to be discretizable and the discrete bins to have relative intensities that correlate to specific values of an observable of interest.

*3.2 Genetic algorithm*

To tackle the problem of data scarcity of blinking emitters, we also develop and implement a genetic algorithm (GA). Genetic algorithms are a class of *unsupervised* optimizers that are particularly apt at solving combinatorial optimization and search problems.[41,42] They are rooted on the idea of generating solutions from an initial pool of guesses that are progressively improved and selected out through evolution-inspired strategies. Operatively, GAs rank *individuals* (i.e., prototype solutions) from a *population*, based on a *fitness function* that measures how well each individual solves the target problem. From the current *generation* of solutions, the fittest individuals—i.e., the best-performing solutions—are then subjected to *crossover* and *mutation* to spawn the next generation of individuals. Briefly, the crossover process takes two (sometimes more) well-performing solutions (*parents*)—as determined by the fitness function—and mixes the values of their parameters to produce a new solution (*child*). Conversely, mutation takes a solution and randomly changes some of its parameters to create a new one (*mutant*); the goal is to allow the algorithm to explore a wider region of the solution space and avoid converging prematurely to local—rather than global—optima. The processes of fitness evaluation and breeding continue through multiple generations until a desired level of accuracy is reached and the fittest individual (solution) is selected.

In our implementation, we start from a single time trace with values of $\langle \tau_{on} \rangle$ and $\langle \tau_{off} \rangle$ to be determined. From the time trace, the two histograms of the on and off occurrences are extracted as per usual (figures 2c–f) and analyzed separately. From each histogram individuals are created. An individual is, simply, a random subset of the histogram data containing (time-bin, occurrences) pairs. These individuals constitute the starting generation. It is worth emphasizing that our GA is unsupervised which means that it runs without requiring a set of known training data (unlike the MFR algorithm of § 3.1); our only input is the time trace of unknown dynamics, which we assume exhibits switching probabilities that follow exponential distributions (cf. § 2.3, hypothesis *I*).

From this starting generation the selection process begins. Each individual is ranked through a fitness function and directly compared to the others and to the conditions for algorithm termination. In our case, the ranking of the individuals is carried out using *silhouette* scoring,[43] applied to *K-means++* clustering.[44] K-means++ takes the (time-bin, occurrences) pairs of an individual and organizes them in $k$ clusters/groups, each identified by its center point, or centroid. The pairs are assigned using Euclidean distance to the closest centroid such that the sum of the squared distance between each point and the centroid is minimized. Briefly, this corresponds to minimizing the potential function:

$$\phi = \sum_{x \in \chi} \min_{c \in C} \|x - c\|^2 \qquad (7)$$

for $k$ centers $C = \{c_1, c_2, \ldots, c_k\}$ and $n$ data points $\chi = \{x_1, x_2, \ldots, x_n\}$ where, again, each data point $x_i$ is a (time-bin, occurrences) pair in our case. Since we want to rank individuals against each other we also need a metric to evaluate this clustering. The silhouette value is a convenient choice for this, as it measures how similar each object—i.e., each (time-bin, occurrences) pair—is to the cluster it has been assigned to, compared to the other clusters. The silhouette score, $s(i)$, is a numeric value that ranges from −1 to +1, where a value close to +1 indicates that the object is well matched to its own cluster and poorly matched to neighboring clusters. Specifically, for the $i$-th data point we define the silhouette score as:[43]

$$s(i) = \begin{cases} 1 - a(i)/b(i), & \text{if } a(i) < b(i) \\ 0, & \text{if } a(i) = b(i) \\ b(i)/a(i) - 1, & \text{if } a(i) > b(i) \end{cases} \quad (8)$$

Where $a(i)$ is the mean distance (e.g., Euclidean distance) between the $i$-th data point and all other data points in the same cluster, and $b(i)$ is the smallest mean distance of the $i$-th point to all points in any other cluster (i.e., in any cluster which the $i$-th point does not belong to).

At the outset, the user defines a strict range for $\langle \tau_{on/off} \rangle$ (e.g., 1–100 ms). Following the loading of the dataset, a heuristic estimate for $\langle \tau_{on/off} \rangle$ is derived based on: *i)* the total observation time of the system, *ii)* the longest observed event, *iii)* the event with the highest frequency and *iv)* the user-defined range. This heuristic helps constrain the solution space to a likely range around the initial estimate. The algorithm executes clustering on the (time-bin, occurrences) pairs of each individual employing a strategy based on K-Means++ (Smartcore Crate[45]), iteratively refining the clustering of groups of randomly sampled points until reaching a predetermined threshold. This threshold indicates that, on iteration, the number of points that undergo cluster reassignment and point movement within clusters adhere to defined tolerances. Fitness is always evaluated in pairs between two groups of $k$ clusters. The group with the highest silhouette score is retained, and the other is dropped. If neither score exceeds the threshold, the top scorer is autonomously replicated (it is cloned, and the clones randomly exchange points), resulting in two offspring that subsequently undergo mutation. Mutations help traverse the solution space, which is particularly useful when $\langle \tau_{on/off} \rangle$ lacks data representation. Increased mutation rates may assist convergence in such scenarios. After mutation the fitness function re-clusters each offspring, initiating the process anew. Should one or both scores exceed the threshold silhouette score the process terminates, returning the most tightly grouped cluster from the top scorer.

Note that, initially, the user estimates the optimal number of clusters, $k$, for the K-means++ algorithm. If the initial clustering does not meet the required level of clustering, the algorithm incrementally adjusts the cluster count until suitable clustering is achieved. The cluster count remains fixed for subsequent evaluations, except after substandard clustering attempts. This is one form of elitism, which we leverage in multiple ways—there is also a penalty incurred once a solution has been accepted, based on the distance of the current estimation from previous estimations. The different forms of elitism allow for favorable traits of 'good' solutions to influence subsequent generations without the need for explicit genetic transfer.

The decay parameter $\tau_{on/off}$ is then determined as:

$$\tau_{on/off} = \frac{D_M}{\ln(2) \cdot [\ln(C_{\max}) - \ln(C_M)]} \quad (9)$$

Where $M$ is the median point of a "winning" cluster; $D_M$ is the duration of the median point; $C_M$ is the occurrence count of the median point; $C_{\max}$ is the occurrence count of the point with maximum duration. The $\ln(2)$ accounts for the difference between the median and mean in the exponential decay function.

The algorithm runs until the rolling estimate stabilizes (using a window of antecedent estimates for reference), or a maximum number of iterations is reached. Whereas convergence yields the final estimate, reaching the maximum iteration count triggers a weighted blend of the final estimate with the mean, mode, and median of all previous estimates.

To test the performance of our GA, we use the synthetic data with $\langle \tau_{on} \rangle = 15$ ms and $\langle \tau_{off} \rangle = 45$ ms as we did for the multi-feature regression (MFR) algorithm and the traditional Levenberg-Marquardt (L-M) fit. The results are shown in Figure 3. Similar to the case of the MFR algorithm, our GA is able to determine the values of $\langle \tau_{on/off} \rangle$ with ≥85% accuracy using a ≥10× factor fewer data points than the L-M method. Notably, the GA is also able to predict $\langle \tau_{on/off} \rangle$ with consistently better precision (by a factor ≥20–40×)—even compared to the MFR algorithm. This is indicated by the small shading (i.e., smaller uncertainty) in the prediction graphs (figures 3a, 3b), and it is also highlighted in the heatmap data (figures 3c, 3d).

We remark that GAs and traditional optimization methods vary significantly in their approaches. GAs strike a balance between exploring new areas of the solution space and exploiting the properties of the local solution space, making them effective in addressing problems characterized by discrete or nonlinear constraints, or those that exhibit discontinuities. A notable benefit of GAs is their capacity to operate without relying on derivatives or assuming differentiability, distinguishing them from most traditional optimization methods. Another prominent characteristic of GAs is their robustness against variations in initial conditions and data. Traditional optimization methods can be highly sensitive to these fluctuations, which may lead to suboptimal results—especially in problems like the one addressed in this work where data is scarce. These features are the main reasons why our GA algorithm is able to significantly outperform a standard L-M fitting method, since the latter suffers greatly when trying to extract the fitting parameters from low frequency events (e.g. low occurrences in the long-time bins).

However, it is crucial to underscore that GAs are computationally expensive. Due to their dependence on population-based search strategies, GAs necessitate an appreciable number of function calls for fitness evaluation and genetic operations. This computational overhead can be limiting, especially when dealing with objective functions that are already computationally intensive. Our implementation can be resource-hungry when working with 'long' observation times (≳30 s), and the time-to-convergence increases significantly (e.g., from ~5 min for a time trace of 2 s to ~45 min for a time trace of 50 s).

### 4. Conclusion

To conclude, we have proposed two approaches based on machine learning (ML) to characterize quantitatively the photodynamics of blinking emitters. The two approaches are a multi-feature regression (MFR) algorithm and a genetic algorithm (GA). Through the employment of synthetic, emulated data, we demonstrate that the algorithms can extract the switching rates of blinking emitters with an accuracy ≥85%, and with ≥10× less data and ≥20× higher precision than traditional statistical inference methods such as Levenberg-Marquardt fitting. As a result, our algorithms effectively extend the range of surveyable blinking and trapping dynamics to include those that would be otherwise beyond investigation due to data scarcity. This has potential positive ramifications towards both gaining a better understanding of photoblinking, and devising ad-hoc strategies to either mitigate or harness the phenomenon in applications based on quantum emitters.

### 5. Data Availability

The data and code used in this study are available upon request. Interested researchers may contact C.B. for access to them.

## 6. Contribution and Acknowledgements
Theoretical modeling, simulation and data analysis was performed by both G.L. and C.B, who co-contribute to the writing of the manuscript.

The Natural Sciences and Engineering Research Council of Canada (via GPIN-2021-03059 and DGECR-2021-00234) and the Canada Foundation for Innovation (via John R. Evans Leaders Fund #41173) are gratefully acknowledged.